\documentclass[prl,twocolumn,aps]{revtex4}
\usepackage{graphicx}
\begin{document}
\title{Thermalization of magnetically trapped metastable helium}
\author{A. Browaeys, A. Robert, O. Sirjean, J. Poupard, S. Nowak, D. Boiron, C. I. Westbrook and A. Aspect}
\affiliation{Laboratoire Charles Fabry de l'Institut d'Optique,
UMR 8501 du CNRS, B.P. 147, F-91403 ORSAY CEDEX, FRANCE}
\begin{abstract}
We have observed thermalization by elastic collisions of
magnetically trapped metastable helium atoms. Our method directly
samples the reconstruction of a thermal energy distribution after
the application of an RF knife. The relaxation time of our sample
towards equilibrium gives an elastic collision rate constant of
$\alpha \sim 5\times 10^{-9}\textrm{cm}^3$/s at a temperature of 1
mK. This value is close to the unitarity limit.
\end{abstract}
 \maketitle
Bose-Einstein condensation (BEC) of dilute atomic vapors has been
observed in Rb \cite{Anderson:95a}, Na \cite{Davis:95a}, Li
\cite{Hulet} and H \cite{Kleppner}. Atoms in these gases are in
their electronic ground state. Metastable helium in the
2~$^3$S$_1$ state (He$^*$), which has long been of interest to the
laser cooling community, is by contrast in a state 20 eV above the
ground state. This situation presents new possibilities for the
study of cold dilute atomic gases. First, the large internal
energy permits efficient detection by ionization of other atoms
and surfaces: it is possible to study very small samples. Second,
Penning ionization by both the background gas and between trapped
atoms offers a high time resolution monitor of the number and
density of trapped atoms. Third, the possibility of using the
large internal energy of He* for atomic lithography has already
been demonstrated \cite{Konstanz}, and this application as well as
atom holography \cite{holo} may benefit from highly coherent
sources. Finally, much theoretical work has already been devoted
to estimation of the elastic collision cross sections on the one
hand and Penning ionizing rates on the other
\cite{Shlyapnikov:94a,venturi:99}. Experiments such as the one
reported here can test this work.

BEC is achieved in dilute gases by evaporative cooling of a
magnetically trapped sample \cite{Hess}. In He$^*$, it is hampered
by the fact that in a magneto-optical trap, the typical starting
point of magnetic trapping, the achievable atomic density is
limited by a large light-assisted Penning ionization rate
\cite{Bardou:92a,Mastwijk:98a,Tol:99a,Kumakura:99a,Browaeys:99a,Pereira}.
On the other hand, the scattering length for low energy elastic
collisions is predicted to be quite large, and the Penning
ionization rate highly suppressed in a spin polarized sample
\cite{Shlyapnikov:94a,venturi:99}. He* in a magnetic trap
necessarily constitutes a spin polarized sample and experiments
have already demonstrated a suppression of more than one order of
magnitude \cite{Hill:72,Nowak:00,Herschbach:00}. If the
theoretical estimates are right, efficient evaporative cooling may
still be possible in spite of the low initial trap density. We
report here the observation of the thermalization of He$^*$ due to
elastic collisions which appears to roughly bear out the
predictions.

To perform a thermalization experiment, a trapped cloud is
deliberately placed out of equilibrium and its relaxation due to
the elastic collisions between trapped particles is observed.
Usually the observations are made by imaging the spatial
distribution as a function of time
\cite{Monroe:93,Davis:95b,Arndt:97}. In our experiment the
relaxation is observed in the energy distribution of the atoms in
the magnetic trap. First this distribution is truncated above
$E_{\text{RF}}=h\nu$ by a radio-frequency pulse (or RF knife) of
frequency $\nu$. The cloud rethermalizes by elastic collisions and
the population of the states of energy higher than $E_{\text{RF}}$
increases from zero; for large times compared to the
thermalization time $\tau_{\text{th}}$ the distribution reaches a
thermal distribution \cite{Snoke:89}. With the help of an
analytical model and numerical simulations, we deduce
$\tau_{\text{th}}$ from the time dependence of the number of atoms
with energy above $E_{\text{RF}}$. We measure this time dependence
by applying a second RF knife after a delay time $t$, and with a
frequency slightly above that of the first one. Our model also
allows us to relate $\tau_{\text{th}}$ to the elastic collision
rate per atom in the trap.

Much of our setup has been described previously
\cite{Browaeys:99a,Nowak:00}. Briefly, we use a $\textrm{LN}_2$
cooled DC discharge source to produce a beam of metastable He
atoms. The beam is slowed down to $\sim 100$~m/s using Zeeman
slowing and loads a magneto-optical trap. Typically, $3\times
10^8$ atoms are trapped at a peak density of $3\times
10^9$~at/cm$^3$, limited by light-induced Penning ionization. The
temperature of the cloud is about 1~mK and the cloud is roughly
spherical with an RMS size of 2.5~mm. We then apply a 5~ms Doppler
molasses to cool the atoms down to 300~$\mu$K. This is achieved by
switching off the magnetic field, decreasing the detuning close to
resonance and lowering the intensity to $10\ \%$ of its value in
the MOT. An optical pumping step allows us to trap up to
$1.5\times 10^8$ atoms in a Ioffe-Pritchard trap. We use a
"cloverleaf" configuration \cite{Mewes:96} with $B'=85$~G/cm,
$B''= 25$~G/cm$^2$ and a bias field $B_0 =200$~G. The two sets of
coils are outside the vacuum, separated by 4~cm. After lowering
the bias field to 4~G, the temperature of the compressed atomic
sample reaches 1~mK. The lifetime of the trap is 60~s.

\begin{figure*}[t]
\begin{center}
\includegraphics[height=5cm]{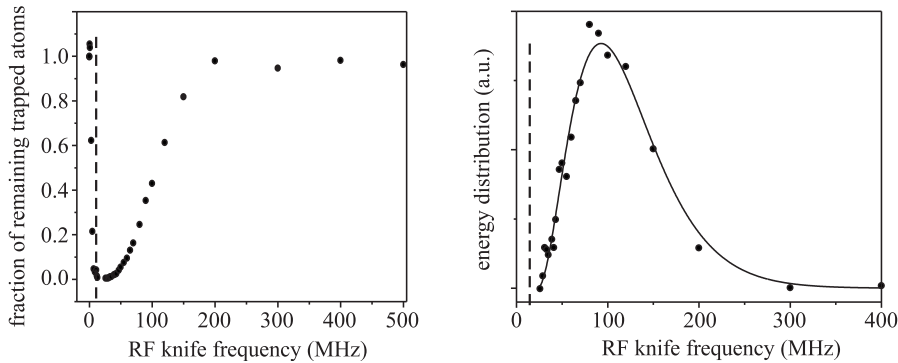}
\end{center}
\caption{RF spectrum of atoms in the magnetic trap.  a) Fraction
of remaining trapped atoms after the RF pulse as a function of the
RF frequency $\nu$. b) Derivative of these data, i.e. the energy
distribution in the magnetic trap. The solid line is the
prediction for a cloud at a temperature of 1.1~mK, the temperature
measured by time of flight (TOF). The dashed line indicates the
frequency corresponding to the bias field.
} \label{figspectro}
\end{figure*}

We use a 2 stage microchannel plate (MCP) to detect the atoms. The
MCP is placed 5~cm below the trapping region and has an active
area of 1.4~cm diameter. Two grids above the MCP allow us to repel
all charged particles and detect only the He*. After turning off
the magnetic trap, the MCP signal corresponds to a time of flight
spectrum (TOF) which gives the temperature of the atoms. The area
of this spectrum is proportional to the number of atoms in the
trap at the time it was turned off. The collection and detection
efficiency of the MCP varies by roughly a factor of two depending
on the magnetic field configuration we use, and so one must take
care to only use data corresponding to the same magnetic field
when making comparisons. We also use the MCP to monitor the atoms
falling out of the trap while applying an RF knife. The area of
the MCP signal in this case measures the number of atoms with an
energy above that of the RF knife. Finally, when we bias the grids
so as to attract positive ions, the MCP signal can be used to
observe the products of Penning ionization with the background gas
while the trap is on. This signal is proportional to the number of
trapped atoms. We observe an exponential decay, indicating that
two body loss (He* + He*) is negligible.

Two parallel coils in the vacuum system produce an RF magnetic
field perpendicular to the bias field and constitutes the RF
knife. To understand the effect of the RF knife on the trapped
cloud and to assure that our sample is at thermal equilibrium, we
first performed an RF spectroscopy measurement of the energy of
the atoms in the trap \cite{Pritchard:88}. We apply an RF pulse at
a frequency $h\nu$ which changes the Zeeman sublevel of the atoms
from the trapped $M = +1$ state to $M = 0$. The duration of the
knife is 3~s, which is necessary to expel all the atoms with
energy above $h\nu$ over the entire range which we explore. We
then turn off the magnetic trap to measure the number of remaining
atoms. Observation of the atoms falling onto the MCP during the RF
knife shows that the flux of atoms expelled is negligible at the
end of the pulse. An example of the RF spectrum is shown in
Fig.~\ref{figspectro}a). The derivative of the data gives the
energy distribution. In Fig.~\ref{figspectro}b) we compare this
distribution with a thermal one at 1.1~mK, the temperature
measured by an independent TOF measurement. We conclude that our
atomic sample is close to thermal equilibrium.

\begin{figure}
\begin{center}
\includegraphics[height=5cm]{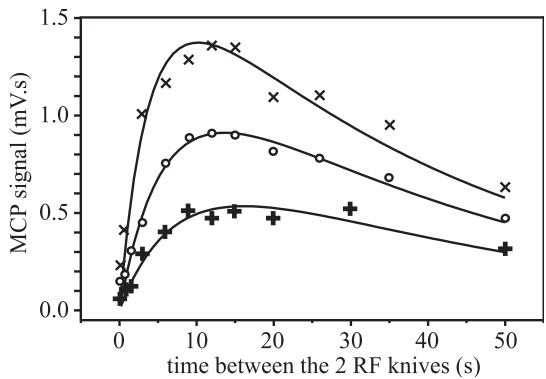}
\end{center}
\caption{
%Relaxation of the trap after the truncation of the
%energy distribution by an RF pulse of 1 sec.
Integrated MCP signal during the RF probe pulse as a function of
the delay between the truncation and probe pulses.  The three
curves correspond to $5\times10^7$, $7\times10^7$ and
$10\times10^7$ atoms in the trap, varied by changing the power in
the Zeeman slowing laser.
%The number of atoms above the
%second RF knife represents at the most a few percent of the number
%of trapped atoms.
 The lifetime of the trap is $38\pm 4$~s, and the
temperature is $0.9\pm 0.1$~mK.} \label{fig3courbes}
\end{figure}

We begin the thermalization experiment with a 2~s RF knife of
frequency $\nu_1 = 135$~MHz (corresponding to
$\eta=(h\nu-2\mu_{\text{B}}B_0)/k_{\text{B}}T \sim 6$). Next we
measure the number of atoms falling onto the MCP during a second
RF knife at a slightly higher frequency ($\nu_2 = 138$~MHz) and
delayed by a time $t$. Assuming that the angular distribution of
the atoms expelled by the second RF knife is constant during the
thermalization process, the MCP signal is proportional to the
number of expelled atoms. Plots of the number of expelled atoms as
a function of $t$ are shown in Fig.~\ref{fig3courbes} for samples
having different numbers of atoms but the same temperature to
within $10\%$. Fig.~\ref{fig3courbes} shows that the number of
atoms above the RF knife increases rapidly and then falls again
with a time constant close to the trap lifetime as atoms are lost.
If the initial increase is indeed due to thermalizing collisions,
the initial slope of each curve should be proportional to the
square of the number of atoms. Our data roughly confirm this
dependence.

To be more quantitative, and to determine the thermalization time
$\tau_{\text{th}}$, we use a model based on the Boltzmann equation
under the sufficient ergodicity hypothesis and inspired by
\cite{Walraven:96}. We divide the sample into two energy regions,
$\mathcal{E}_-$ and $\mathcal{E}_+$, with energies below and above
$\eta k_{\text{B}}T$ respectively and denote by $N_-$ and $N_+$
the number of atoms belonging to the two regions. We assume that
$\eta \gg 1$. Immediately after truncation, $N_+=0$, and we seek a
differential equation governing the time dependence of $N_+$.
Since $\eta \gg 1$, we only take into account collisions of the
type $(\mathcal{E}_-)+(\mathcal{E}_-)
\leftrightarrow(\mathcal{E}_-) +(\mathcal{E}_+)$, and neglect all
collisions involving two atoms in $\mathcal{E}_+$ in either the
final or initial state. The corresponding flux $\dot{N}_+$ is thus
of the form
\begin{equation}\dot{N}_+ = \Delta_1 {N_-}^2-\Delta_2
N_-N_+.
\end{equation}
The coefficients $\Delta_1$ and $\Delta_2$ are calculated using
Boltzmann equation \cite{db}. In particular $\Delta_1 {N_-}$ is
exactly the evaporation rate in an evaporative cooling process
\cite{Walraven:96}. If we make the further approximations that the
atoms in $\mathcal{E}_-$ and $\mathcal{E}_+$ have thermal
distributions \cite{explication}, neglect variations of the
temperature during thermalization and assume that the collision
cross section $\sigma$ is independent of velocity, $\Delta_1$ and
$\Delta_2$ are analytic functions of the trap parameters, atomic
mass $m$, $\sigma$, $\eta$ and $\mu_{\text{B}}B_0/k_{\text{B}}T$.
This latter parameter appears because our trap cannot be
approximated by an harmonic trap; we use the semi-linear form
\cite{Walraven:96}. It is straightforward to take into account the
finite lifetime $\tau$ of the atomic sample since
$N_-(t)+N_+(t)=N_-(0)\exp{(-t/\tau)}$. The solution of the
resulting differential equation is :
\begin{equation}\label{soleqthermal}
N_+(t)=N_{\text{th}}e^{-t/\tau}[1+\frac{q}{1-q-\exp{[\frac{\tau}{\tau_{\text{th}}}
(1-e^{-t/\tau}) ]}}]
\end{equation}
where
$\tau_{\text{th}}^{-1}=\frac{\gamma_{\text{el}}}{\sqrt{2}}\frac{q}{1-q}\frac{e^{-\eta}
V_{\text{ev}}}{V_{\text{e}}}$ and $N_{\text{th}}=(1-q)N_-(0)$.
\vskip 6pt

\noindent The elastic collision rate is
$\gamma_{\text{el}}=n\sigma \overline{v}$ with $n$ defined at the
center of the trap and $\overline{v}=4\sqrt{k_{\text{B}}T/\pi m}$.
The quantities $V_{\text{ev}}$, $V_{\text{e}}$ and $q$ are defined
as in \cite{Walraven:96,db}; they are analytic functions of $\eta$
and $\mu_{\text{B}}B_0/k_{\text{B}}T$. The quantity $q$ is the
ratio of the number of atoms below the RF knife to the total for a
thermal distribution (about 0.9 under our conditions), and
$N_{\text{th}}$ is the asymptotic value of $N_+$ for infinite trap
lifetime. Numerical simulations of the energy form of Boltzmann
equation are in good agreement with our model for $\eta
> 10$; for $\eta =6$ the quantity
$\gamma_{\text{el}}\tau_{\text{th}}$ is $1.8$ times larger meaning
that our assumption about the distribution function fails for
small $\eta$ \cite{db}. We take this factor into account in
calculating $\gamma_{\text{el}}$.

To fit the data of Fig.~\ref{fig3courbes} with
eq.~(\ref{soleqthermal}), we fix the lifetime $\tau$ at its
measured value and use $\tau_{\text{th}}$ and $N_{\text{th}}$ as
adjustable parameters. The uncertainty in $\tau_{\text{th}}$ is
estimated by varying the lifetime of the trap within its
uncertainty range and looking at the resulting dispersion in
$\tau_{\text{th}}$. The uncertainty in the number of trapped atoms
is estimated from the dispersion of the TOF area measurements
before and after taking a curve as in Fig.~\ref{fig3courbes}. The
exact value of $q$ has little influence on the fit.

\begin{figure}
\begin{center}
\includegraphics[height=5cm]{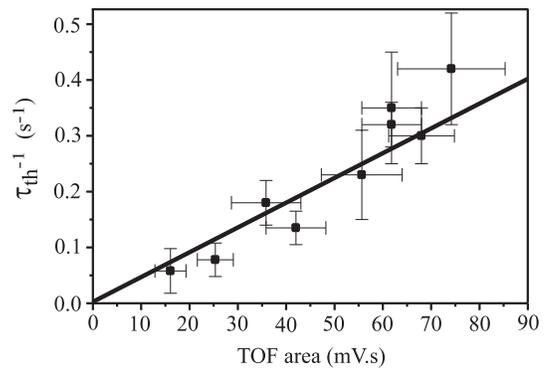}
\end{center}
\caption{Thermalization rate $\tau_{\text{th}}^{-1}$ versus the
area of the corresponding TOF spectrum (proportional to the number
of trapped atoms). The solid line shows a linear fit constrained
to pass through the origin. } \label{figresults}
\end{figure}

We have made several tests to check the consistency of our
results. First, we have checked that the fitted value of
$N_{\text{th}}$ corresponds to the expected fraction of atoms
above the knife for our temperature. Second, Fig.~\ref{figresults}
shows that $\tau_{\text{th}}^{-1}$ is proportional to the number
of trapped atoms, as it must be if the process of refilling of the
upper energy class is due to two body collisions. We can exclude
any effect independent of the number of atoms. The line passing
through the origin uses the slope as a fit parameter and has
$\chi^2= 5$ for 8 degrees of freedom. Third we have done an
additional experiment that confirms the presence of elastic
collisions: in a trap decay rate experiment, {\it in the presence}
of the RF knife, the ion signal exhibits a clear non-exponential
behaviour at short times. This effect can be satisfactorily
interpreted as elastic collisions bringing atoms above the RF
knife and hence allows a measurement of the evaporation rate. This
rate is consistent with the results obtained in our thermalization
experiment. Fourth we have checked that heating cannot explain the
repopulation of the upper energy classes. With the trap
undisturbed, we can place an upper limit on the heating rate of
$25\,\mu$K in 60~s. This limit is two orders of magnitude too low
to explain our data. Lastly, we have performed the thermalization
experiment for different lifetimes of the magnetic trap (20, 40
and 60~s) and found consistent results.

From our data in Fig.~\ref{figresults}, we can deduce an accurate
measurement of the thermalization time; the fit gives
$\tau_{\text{th}}=3.0\pm 0.3$~s for the densest sample. Using the
measured temperature and bias field, this value of
$\tau_{\text{th}}$ leads to $\gamma_{\text{el}}=6\pm 1$~s$^{-1}$;
this result depends on the accuracy of our thermalization model.
To find the rate constant $\alpha=\gamma_{\text{el}}/n$, we must
estimate the density. Since the data show that our sample is close
to thermal equilibrium, we can calculate the volume of the trap
knowing the trap parameters. The absolute measurement of the
number of atoms is performed by measuring the total power absorbed
from a saturating laser beam, similar to \cite{Pereira}. A TOF
area of $75$~mV.s corresponds to $10^{8}$ atoms in the magnetic
trap with an uncertainty of a factor of 2. This leads to $\alpha=
5 \times 10^{-9}$~cm$^3$/s to within a factor 3 at $T=1\pm
0.1$~mK. The ENS-Paris group has obtained a similar result with a
different measurement \cite{Leduc}. The unitarity limit at that
temperature is $\alpha \sim 10^{-8}$~cm$^3$/s. This means that it
is probably not valid to use a constant elastic cross section in
our model and some deviation might appear in the quantity
$\gamma_{\text{el}}\tau_{\text{th}}$. We are currently
investigating refinements to our thermalization model.

The results shown here are very encouraging for evaporative
cooling of He* in search of BEC.\\

%\begin{acknowledgments}
We thank P. Leo and P. Julienne and the ENS helium group for
helpful discussions. This work was partially supported by the EC
under contracts IST-1999-11055 and HPRN-CT-2000-00125, and DGA
grant 99.34.050
%\end{acknowledgments}

%\bibliographystyle{Vbibsty}
%\bibliography{C:/tex/tex/latex/base/biblio}

\begin{thebibliography}{2}
\bibitem{explication}
This hypothesis is legitimate in $\mathcal{E}_-$ because it is
true at $t=0$ and because only few particles will leave this
region; one can find indications in \cite{Snoke:89} that this
hypothesis is not too strong in $\mathcal{E}_+$.
\end{thebibliography}

\end{document}